\documentclass[10pt]{iopart}
\usepackage[utf8]{inputenc}
\usepackage{subcaption}
\usepackage{graphicx} 
\expandafter\let\csname equation*\endcsname\relax

\expandafter\let\csname endequation*\endcsname\relax
\usepackage{amsmath}
\usepackage{amsfonts} 
\usepackage{booktabs}
\usepackage{multirow}
\usepackage{cite}
\usepackage{enumitem}
\usepackage{algorithm}
\usepackage{algorithmic}
\usepackage{pifont}

\usepackage[round,comma,sort&compress]{natbib}

\begin{document}

\title{Brain Network Diffusion-Driven fMRI Connectivity Augmentation for Enhanced Autism Spectrum Disorder Diagnosis}

\author{Haokai Zhao$^1$, Haowei Lou$^1$, Lina Yao$^{1,2}$, Yu Zhang$^{3,4}$}

\address{$^1$ Computer Science Building (K17), Engineering Rd, UNSW Sydney, Kensington NSW 2052}
\address{$^2$ CSIRO's Data61, Level 5/13 Garden St, Eveleigh NSW 2015}
\address{$^3$ Department of Bioengineering, Lehigh University, Bethlehem, PA 18015, USA}
\address{$^4$ Department of Electrical and Computer Engineering, Lehigh University, Bethlehem, PA 18015, USA}
\eads{z5261811@ad.unsw.edu.au, haowei.lou@student.unsw.edu.au, lina.yao@data61.csiro.au, yuzi20@lehigh.edu}
\vspace{10pt}
\begin{indented}
\item[]August 2024
\end{indented}

\begin{abstract}
Functional magnetic resonance imaging (fMRI) is an emerging neuroimaging modality that is commonly modeled as networks of Regions of Interest (ROIs) and their connections, named functional connectivity, for understanding the brain functions and mental disorders. However, due to the high cost of fMRI data acquisition and labeling, the amount of fMRI data is usually small, which largely limits the performance of recognition models. With the rise of generative models, especially diffusion models, the ability to generate realistic samples close to the real data distribution has been widely used for data augmentations. In this work, we present a transformer-based latent diffusion model for functional connectivity generation and demonstrate the effectiveness of the diffusion model as an augmentation tool for fMRI functional connectivity. Furthermore, extended experiments are conducted to provide detailed analysis of the generation quality and interpretations for the learned feature pattern. Our code will be made public upon acceptance.
\end{abstract}

\vspace{2pc}

\submitto{\JNE}

\maketitle

\ioptwocol

\section{Introduction}
Autism Spectrum Disorder (ASD) is a neurodevelopmental deficits that impacts how individuals communicate and interact with others~\citep{lord2018autism}. Common diagnostic practices include medical and neurological examinations, assessments of an individual's cognitive and language abilities, and observation of behavior~\citep{vahia2013diagnostic,lord2018autism}. Among neurological examinations, Functional Magnetic Resonance Imaging (fMRI) has been widely adopted due to its ability to detect changes in blood flow within the brain, providing insights into the neural mechanisms underlying ASD~\citep{ecker2015neuroimaging,muller2018brain}. In practice, fMRI data is used to construct functional connectivity by analyzing blood-oxygen-level-dependent~(BOLD) signals. These signals help examine functional correlations between Regions of Interest~(ROIs) as defined by a brain atlas~\citep{keown2013local,assaf2010abnormal,hull2017resting,han2024early}.

The advent of Machine Learning (ML) brings many ML-based diagnostic frameworks that can automatically diagnose a patient's ASD status by analyzing functional connectivity~\citep{haghighat2022sex,wang2019functional,sadeghian2021feature,yan2019groupinn,li2021braingnn,kan2022brain}. ML-based framework require lots of data to achieve good performance. However, collecting high-quality fMRI data necessitates professional operators, specialized equipment, and dedicated lab environments. This process is both expensive and resource-intensive. The problem of data scarcity limits the performance and generalizability of ML models for ASD diagnosis.

To address the problem of data scarcity, many researchers have tried data augmentation techniques to artificially expand the training dataset. For example, \citep{pei2022data} proposed to add random Gaussian noise on functional connectivity and multiple functional connectivity construction with sliding window along ROI signals as data augmentation to improve the identification of Attention Deficit/Hyperactivity Disorder (ADHD) in fMRI studies. \citep{wang2022contrastive} used sliding windows of ROI signals as data augmentation technique to construct contrastive pairs for contrastive representation learning. 

Building on these traditional methods, deep learning based approaches have pushed the frontier of data augmentation techniques. \citep{qiang2021modeling} introduced RNN based Variational AutoEncoder~(VAE) for ROI signals generation and augmentations to improve ADHD diagnosis. \citep{li2021brainnetgan} used a GAN-based generator to generate functional connectivity for data augmentation of Alzheimer’s disease diagnosis. \citep{zhuang2019fmri} proposed GAN and VAE combined framework that utilize 3-dimensional convolutions to generate high-dimensional brain image tensors for augmenting cognitive and behavioural prediction task. \citep{xia2021learning} provided an analysis of aging process and demonstrate the potential for data augmentation with conditional GAN. \citep{fang2024action} developed an open-source software for fMRI data analysis and augmentation, covering BOLD signal augmentation and brain network augmentation. However, VAE \& GAN are experiencing the problem of posterior collapse~\citep{lucas2019understanding} and mode collapse~\citep{metz2016unrolled} respectively that can affect the generation quality. 

Diffusion methods~\citep{ho2020denoising,rombach2022high}, as a strong and competitive alternative generative model, have been applied to augment various types of medical images. These include micro-array images~\citep{ye2023synthetic}, skin images~\citep{akrout2023diffusion}, lung ultrasound images~\citep{zhang2023diffusion}, 3D brain MRI images~\citep{pinaya2022brain}, and chest X-Ray images exhibiting various abnormalities~\citep{chambon2022adapting,packhauser2023generation}.

Despite these advancements, there is currently no diffusion-based data augmentation method specifically tailored for brain networks in the form of functional connectivity. Adapting diffusion models to generate functional connectivity faces several challenges. First, the non-Euclidean nature of functional connectivity does not meet the prerequisites of convolutional neural networks (CNNs), making standard diffusion models, which are typically designed for image generation, unsuitable. Second, the numeric distribution of functional connectivity does not conform to the standard Gaussian distribution. As such, the mismatch between the signal and noise distributions, where classic diffusion models use standard Gaussian noise, challenges the assumption that input signals and noise follow the same distribution~\citep{ho2020denoising}. 
Finally, the small data size and complexity of functional connectivity patterns make it difficult for the diffusion model to effectively learn condition-specific features, rendering conditioning mechanisms like adaptive layer normalization ineffective.

In this work, we proposed Brain-Net-Diffusion, a pure transformer-based latent diffusion model for fMRI data generation and augmentation in the form of functional connectivity. Developed based on Diffusion Transformer~\citep{peebles2023scalable}, our Brain-Net-Diffusion has three key designs which make it more suitable for functional connectivity generation. Firstly, Brain-Net-Diffusion uses transformers for both latent ROI feature encoding and diffusion denoising. The attention mechanism allows modeling interactions between ROIs all at once. Furthermore, to resolve the problem of distribution gap between encoded latent space and scheduled noise in diffusion process, Brain-Net-Diffusion uses distribution normalization module to ensure the generated connectivity follows the same numerical distribution with the real connectivity. Finally, to tackle the problem of ineffective conditioning during diffusion process, Brain-Net-Diffusion uses an extra constraint on conditional embedding to ensure the effectiveness of conditioning. 

Our contributions can be summarized as following:
\begin{itemize}
    \item We proposed Brain-Net-Diffusion, a transformer-based latent diffusion model specifically designed for functional connectivity modeling and generation.
    \item We introduced an distribution normalization module to tackle the problems of distribution gap between encoded latent space and scheduled noise.
    \item We introduced conditional contrastive loss to address the problem of ineffective conditioning mechanism for adapting diffusion transformer to resting state fMRI functional connectivity.
    \item  We conduct extensive experiments to assess the quality of the generated data and demonstrate that our data augmentation method can effectively improve ML model performance, outperforming 3\% of existing state-of-the-art methods.
\end{itemize}

\section{Brain Network Diffusion Model}

\begin{figure*}[htbp]
    \centering
    \includegraphics[width=0.8\textwidth]{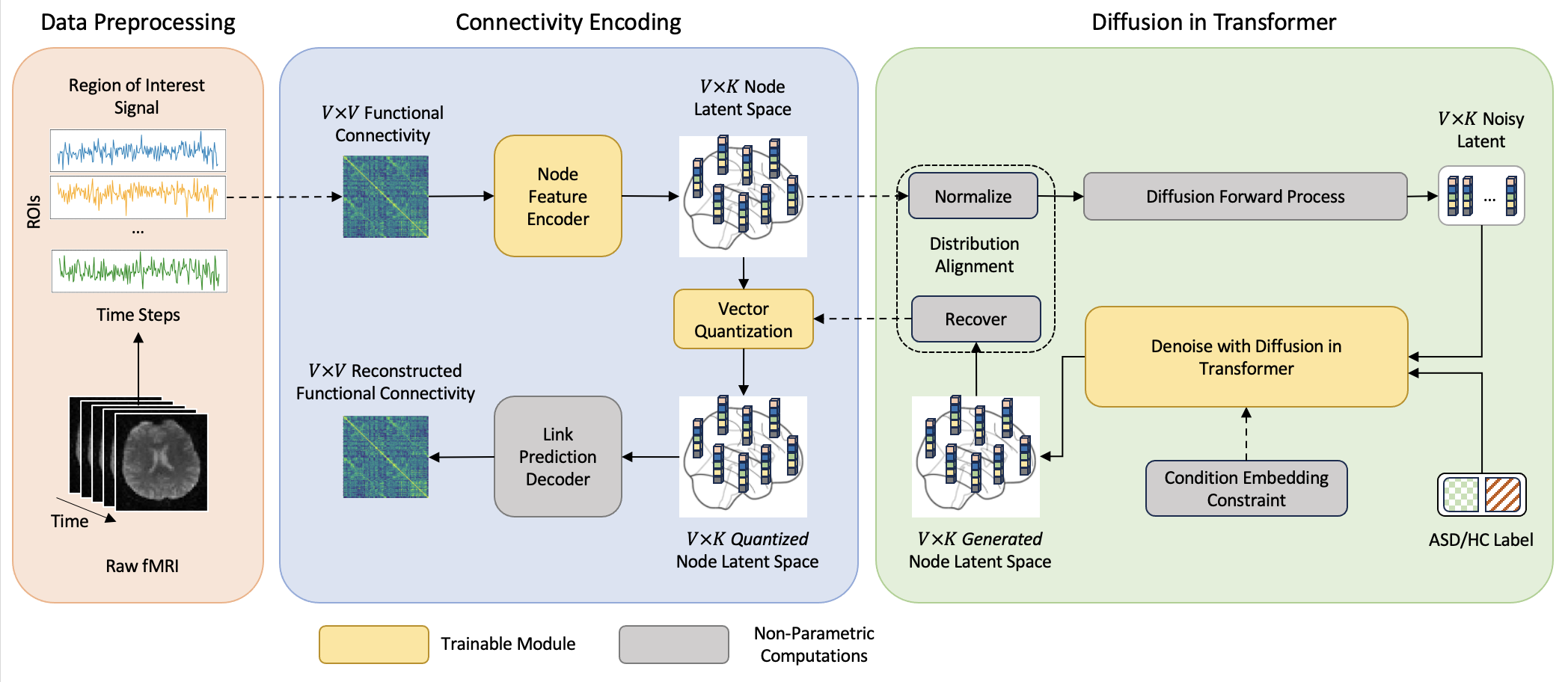}
    \caption{The architecture of Brain-Net-Diffusion framework for functional functional connectivity generation}
    \label{fig:overview}
\end{figure*}

\textbf{Brain-Net-Diffusion} is a diffusion-based generative model designed to generate in-distribution and diverse functional connectivity matrices. These generated matrices are intended to address the issue of data scarcity and enhance the performance of downstream ASD classification models.
Figure \ref{fig:overview} provides an overview of the Brain-Net-Diffusion model. The model is composed of the following three main modules.

\textbf{Latent Connectivity Auto-Encoder}, which encodes functional connectivity into a latent space that contains representations for each ROI. \textbf{Conditional Diffusion Transformer} which generates new latent representations using diffusion model. \textbf{Functional Connectivity Generator} which integrates the Latent ROI Auto-Encoder and Conditional Diffusion Transformer to generate new functional connectivity matrices.

Latent Connectivity Auto-Encoder and Conditional Diffusion Transformer are trained separately, each with distinct loss functions. Further details on these modules and training objectives will be provided in the following sections. 

\begin{figure*}[ht]
    \centering
    \begin{subfigure}[b]{0.32\textwidth}
        \centering
        \includegraphics[width=\textwidth]{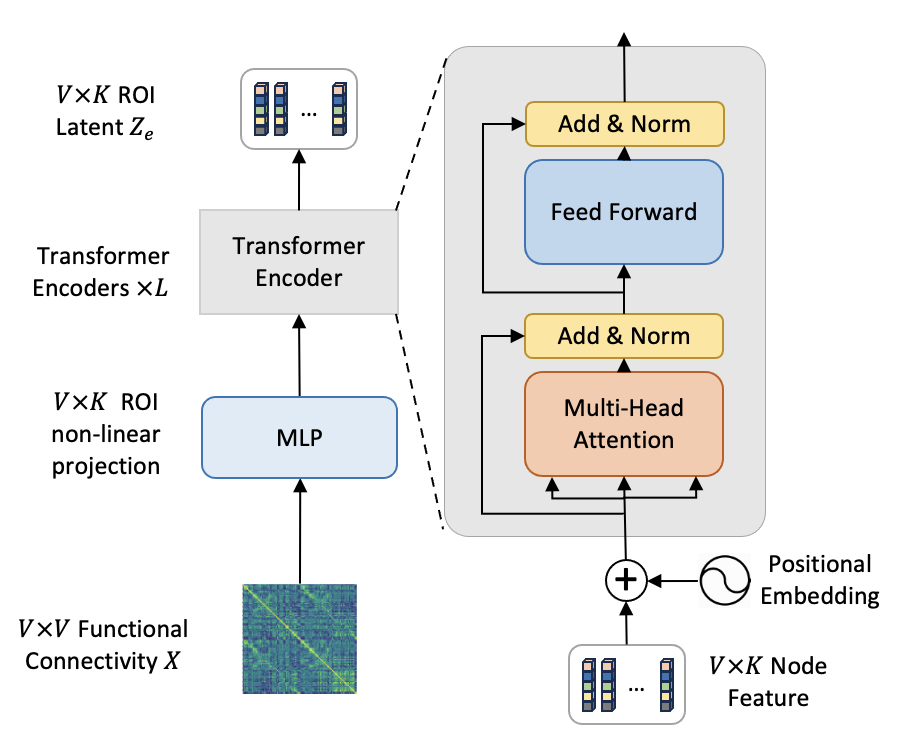}
        \caption{Latent Encoder}
        \label{fig:train_enc}
    \end{subfigure}\hfill
    \begin{subfigure}[b]{0.32\textwidth}
        \centering
        \includegraphics[width=\textwidth]{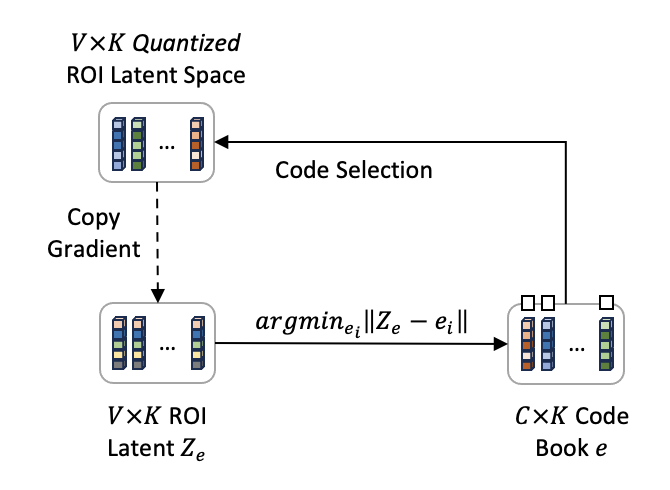}
        \caption{Vector Quantization}
        \label{fig:vq}
    \end{subfigure}\hfill
    \begin{subfigure}[b]{0.32\textwidth}
        \centering
        \includegraphics[width=\textwidth]{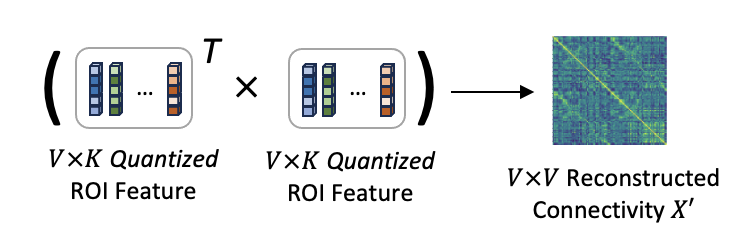}
        \caption{Link Prediction Decoder}
        \label{fig:link_dec}
    \end{subfigure}
    \caption{Transformer-based node auto-encoding with vector quantization}
    \label{fig:node_enc}
\end{figure*}

\subsection{Latent Connectivity Auto-Encoder} \label{roi_autoencoder}
We deployed a Vector Quantized-Variational Autoencoder~(VQ-VAE) as the Latent Connectivity Auto-Encoder to efficiently encode and discretize functional connectivity matrices into a structured latent space.

Figure~\ref{fig:node_enc} illustrates the architecture of the VQ-VAE, which consists of three key components:  (1) Encoder that maps the functional connectivity matrix $X \in \mathbb{R}^{V \times V}$ to a latent representation $Z_e \in \mathbb{R}^{V \times K}$, where $V$ is the number of ROIs and $K$ is the latent feature dimension for each ROI; (2) Vector Quantization (VQ, \citep{van2017neural}), which converts $Z_e$ into a quantized latent representation $Z_q \in \mathbb{R}^{V \times K}$; and a (3) Decoder, that reconstructs the functional connectivity matrix $X' \in \mathbb{R}^{V \times V}$ from $Z_q$.

\subsubsection{Latent Encoder} \label{encoder}
Given a functional connectivity matrix $X \in \mathbb{R}^{V\times V}$, where each row of $X$ represents the correlation of each ROI with all other ROIs, we projected $X$ to a $K$-dimensional feature space using a linear transformation followed by a ReLU activation function. This process produced transformed ROI features~$H \in \mathbb{R}^{V\times K}$, where each row represents the transformed feature of an ROI. We then passed $H$ through $L$ layers of a Transformer encoder \citep{vaswani2017attention}, where each ROI's transformed feature served as a token. We leveraged the capability of multi-head self-attention to dynamically weigh the importance of each ROI during the encoding process, producing latent features~$Z_e \in \mathbb{R}^{V \times K}$. More details on the encoding process can be found in Figure~\ref{fig:train_enc}.

\subsubsection{Vector Quantization} \label{vq}
Following the approach of the latent diffusion model \citep{rombach2022high}, we used VQ to discretize the latent features and address issues such as "posterior collapse" \citep{van2017neural}. 

A codebook $e \in \mathbb{R}^{C \times K}$ is initialized with a set of learnable representation vectors, where $C$ is the size of the codebook, and $K$ is the dimensionality of the codebook vectors. For each latent feature vector $z_e \in \mathbb{R}^K$ in the latent space $Z_e \in \mathbb{R}^{V \times K}$, the corresponding quantized latent feature $z_q \in \mathbb{R}^{K}$ is determined by finding the nearest neighbor in the codebook, as shown in Equation~\ref{qe:quantize}.
\begin{equation}\label{qe:quantize}
    z_q = e_{i}, \, \text{where} \, i = \mathbf{argmin}_{j}\left\|z_e - e_j \right\|_{2}
\end{equation}

During training, the codebook vectors are iteratively updated to minimize the reconstruction loss. This is achieved by adjusting the codebook vectors to be closer to the input vectors assigned to them, typically using the $L2$ loss to measure the difference. The VQ operation is illustrated in Figure \ref{fig:vq}.

\subsubsection{Functional Connectivity Decoder} \label{decoder}
To reconstruct the functional connectivity from the quantized latent features~$Z_q$, we adapted the matrix multiplication decoding approach used in link prediction tasks \citep{kipf2016variational, salha2019keep} as our decoder. The output functional connectivity matrix~$X'$ is reconstructed by computing the dot product~($\odot$) of the quantized latent features, i.e., $X'_{ij} = z_{q(i)} \odot z_{q(j)}$, $X' = Z_q^T \times Z_q$.

\subsubsection{Training Objective}
Followed \citep{van2017neural}, we combined VQ, reconstruction and commitment loss to train our Latent Connectivity Auto-Encoder, as shown in Equation~\ref{eq:encode_loss}: 
\begin{equation}\label{eq:encode_loss}
    \mathcal{L}_{encode} = \mathcal{L}_{VQ} + \mathcal{L}_{reconstruct} + \mathcal{L}_{commitment}
\end{equation}
\begin{align*}
    \mathcal{L}_{VQ} &= \|\mathrm{sg}[Z_e] - e\|_{2}^{2} \\
    \mathcal{L}_{reconstruct} &= \log p(X \mid Z_q) \\
    \mathcal{L}_{commitment} &= \|Z_e - \mathrm{sg}[e]\|_{2}^{2}
\end{align*}

where $Z_e \in \mathbb{R}^{V \times K}$ denotes the latent features produced by the encoder, $e \in \mathbb{R}^{C \times K}$ refers to the codebook containing $C$ learnable vectors, $X \in \mathbb{R}^{V \times V}$ is the original functional connectivity matrix, and $Z_q \in \mathbb{R}^{V \times K}$ represents the quantized latent features derived from the codebook, and the operator $\mathrm{sg}[\cdot]$ is the stop-gradient operator, which prevents gradients from flowing through its argument during backpropagation. 

The VQ loss $\mathcal{L}_{VQ}$ adjust the codebook vectors \(e\) towards the encoder outputs $Z_e$, the reconstruction loss $\mathcal{L}_{reconstruct}$ is used to optimize the encoder, ensuring that the reconstructed matrix $X'$ closely resembles the original functional connectivity matrix $X$, and the commitment loss $\mathcal{L}_{commitment}$ is employed to encourage the encoder outputs to map to the nearest feature in the codebook.

\subsection{Conditional Diffusion Transformer} \label{latent_diffusion}
Once the latent Connectivity Auto-Encoder is trained, it allows us to encode functional connectivity into latent representations $Z_e$. We transformed it into a Gaussian distribution $\hat{Z}_e$ by applying z-score normalization to $Z_e$. We then train a diffusion model over $\hat{Z}_e$, with the model conditioned on the subject's status~$c$ (such as ASD or HC). 

\subsubsection{Forward Diffusion Process}
The forward diffusion process gradually adds noise to $Z_0 = \hat{Z_e}$ at each timestep $t$. This process is mathematically described by the following transition distribution: 
\begin{equation}
    q(Z_t | Z_0) = \mathcal{N}(Z_t; \sqrt{\bar{\alpha}_t} Z_0, (1 - \bar{\alpha}_t) \mathbf{I})
\end{equation}
where $\mathcal{N}$ represents a normal distribution, $\bar{\alpha}_t$ is a time-dependent scaling factor, and $I$ is identity matrix. The parameters $\bar{\alpha}_t$ control the amount of added noise in the process.

To facilitate the generation of $Z_t$ at a specific timestep $t$, we reparameterize the process as:
\begin{equation}
Z_t = \sqrt{\bar{\alpha}_t} Z_0 + \sqrt{1 - \bar{\alpha}_t} \epsilon_t, \quad \text{where} \quad \epsilon_t \sim \mathcal{N}(0, \mathbf{I})
\end{equation}
where $\epsilon_t$ is sampled from a standard normal distribution~$\mathcal{N}(0, \mathbf{I})$.

\subsubsection{Reverse Diffusion Process} \label{reverse_diffusion}
The reverse diffusion process is designed to systematically denoise $Z_t$ and recover the original $Z_0$
 , which corresponds to the normalized latent feature $\hat{Z_e}$.  This process is modelled as a series of conditional distributions, where the transition from 
$Z_t$  to $Z_{t-1}$ at each timestep is given by:
\begin{equation}
p_\theta(Z_{t-1} | Z_t) = \mathcal{N}(\mu_\theta(Z_t), \Sigma_\theta(Z_t))
\end{equation}
where $\mathcal{N}$ denotes a normal distribution, with the mean  \( \mu_\theta(Z_t) \) and covariance \( \Sigma_\theta(Z_t) \) predicted by a transformer-based noise prediction network, which is parameterized by the learnable parameters $\theta$.

\begin{figure}
    \centering
    \includegraphics[width=0.2\textwidth]{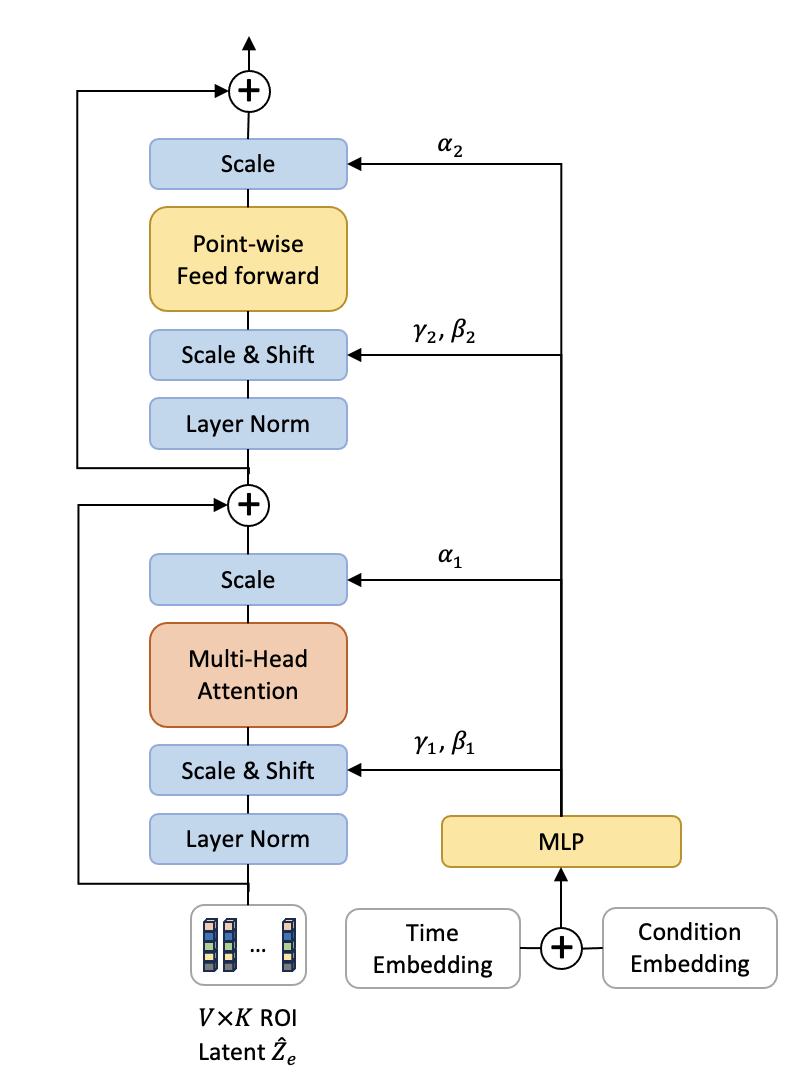}
    \caption{Diffusion Transformer Block}
    \label{fig:dit_block}
\end{figure}

We used the same architecture as the Diffusion Transformer \citep{peebles2023scalable} for our noise prediction network $\epsilon_\theta$ in the reverse diffusion process, as shown in Figure \ref{fig:dit_block}. It consists of a series of transformer blocks that predict the noise present in \( Z_t \). To effectively incorporate conditioning information, we use adaptive layer normalization (adaLN), where the normalization parameters are dynamically adjusted based on the conditioning input:
\begin{equation}
\text{adaLN}(Z_t, t, c) = \gamma(t, c) \cdot \frac{Z_t - \mu(Z_t)}{\sigma(Z_t)} + \beta(t, c)
\label{eq:adaln}
\end{equation}
Here, \( \gamma(t, c) \) and \( \beta(t, c) \) are learned functions that depend on the conditioning variables \( t \) (time step) and \( c \) (class label).

\subsubsection{Training Objective}
The original diffusion transformer noise predictor is trained by minimizing the L2 loss between the predicted noise \( \epsilon_\theta(Z_t, t, c) \) and the true noise \( \epsilon_t \):
\begin{equation}
\mathcal{L}_{noise} = \mathbb{E}_{Z_0, \epsilon, t} \left[ \left\| \epsilon - \epsilon_\theta(Z_t, t, c) \right\|_2^2 \right]
\end{equation}

where \( Z_0 \) (which equals \( \hat{Z_e} \)) is the initial normalized latent feature, \( Z_t \) is the noisy latent feature at timestep \( t \), \(\epsilon\) is the Gaussian noise, and \(\epsilon_\theta(Z_t, t, c)\) is the noise prediction network predicted noise at timestep \( t \).

To enhance the distinctiveness of class embedding and ensure the effectiveness of both time and condition embedding, we introduced \textit{condition contrastive loss}, which contains two additional loss terms:

\textit{Maximizing the distances between each label embedding}: 
We start with an embedding table \(E \in \mathbb{R}^{C \times d}\), where \(C\) is the number of classes, and \(d\)  is the dimension of each embedding. The pairwise distance matrix  \(D \in \mathbb{R}^{C \times C}\) captures the distances between every pair of embedding $e_i$ and $e_j$ using the Euclidean distance:
\begin{equation}
D_{ij} = \| \mathbf{e}_i - \mathbf{e}_j \|_2
\end{equation}
To ensure that each class has a distinct embedding, the first loss term focuses on maximizing the distances between different class embedding (excluding the distance of an embedding to itself):
\begin{equation}
\mathcal{L}_{dist} = \exp\left( -\sum_{i=1}^{C} \sum_{j=1, j \neq i}^{C} D_{ij} \right)
\end{equation}

\textit{Balancing time and condition embedding}: 
To make sure both time and condition embedding are equally effective, we introduce a second loss term. This term minimizes the difference in mean of element-wise absolute value between the time embedding and the condition embedding:

\begin{equation}
\mathcal{L}_{scale} = \left| \frac{1}{n} \sum^{n}_{i=1} \left| y_i \right| - \frac{1}{n} \sum^{n}_{i=1} \left| t_i \right| \right|
\end{equation}

where $t$ is the time embedding, $y$ is the condition embedding and $n$ is the dimension for both embedding.

The overall loss for the noise prediction network combines the original noise prediction loss with these additional terms:
\begin{equation} \label{eq:diffusion_loss}
\mathcal{L}_{\epsilon_\theta} = \mathcal{L}_{noise} + \mathcal{L}_{dist} + \mathcal{L}_{scale}
\end{equation}
where $\mathcal{L}_{noise}$ minimize the sampled and predicted loss, \(\mathcal{L}_{dist}\) maximizes the pairwise distances between class embedding, and \(\mathcal{L}_{scale}\) balances the numerical scales of the embedding.

\subsubsection{Distribution Normalization} \label{dist_align}
We introduced the distribution normalization module to address the numerical distribution gap between the scheduled noise in the diffusion process and the encoded ROI latent features. The distribution normalization module normalizes the ROI features to a standard Gaussian distribution with Z-score normalization before the diffusion training and generation processes. The encoded latent representation \( Z_e \) is first normalized to \( \hat{Z}_e \), which is then used during the diffusion process.

After the reverse diffusion process generates the new ROI latent representation \( \hat{Z}_g \), it is de-normalized to the original latent space, transforming it into \( Z_g \). This ensures that \( Z_g \) matches the distribution of the original encoded latent feature \( Z_e \). 

\begin{figure}
    \centering
    \includegraphics[width=0.2\textwidth]{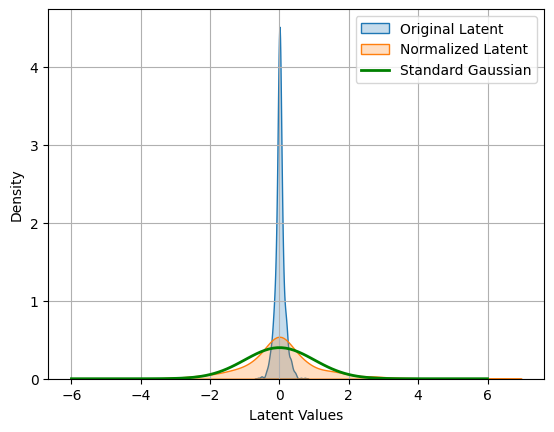}
    \caption{Numerical distribution of latent space}
    \label{fig:latent_dist}
\end{figure}

Specifically, for each data batch that was inputed into the forward diffusion process, we applied z-score normalization to the latent distribution:
\begin{equation}
\hat{Z}_e = \frac{Z_e - \mu}{\sigma}
\label{eqn:z_norm}
\end{equation}
where \( \mu \) is the batch mean and \( \sigma \) is the batch standard deviation. The diffusion transformer was trained on and generated within the normalized latent space \( \hat{Z}_e \). During the generation process, we compute the batch mean and standard deviation of the original latent features, then de-normalize the generated latent features \( \hat{Z}_g \) back to the original latent space using the reverse of Equation~\ref{eqn:z_norm}. The distribution comparison of the original and normalized latent spaces is shown in Figure \ref{fig:latent_dist}.

\subsection{Functional Connectivity Generator}
The trained VQ-VAE (Sec. \ref{roi_autoencoder}) and the Conditional Diffusion Transformer (Sec. \ref{latent_diffusion}) are deployed together for new connectivity generation.
The generation process begins with an initial noise sample \( Z_{t_{\text{max}}} \sim \mathcal{N}(0, \mathbf{I}) \) and a dsired condition $c$ for the generated sample. This noisy representation is then processed by the reverse diffusion process with the trained noise predition network \(\epsilon_\theta(Z_t, t, c)\). At each step, the model predicts and removes the noise:
\begin{equation}
Z_{t-1} \sim p_\theta(Z_{t-1} | Z_t, c)
\end{equation}
iteratively refining $Z_T$ over $T$ steps until the noise is fully removed. It produces a latent feature $Z_G$ that accurately reflects the desired condition.

Next, $Z_g$ is de-normalized using the reverse Z-score function to ensure it matches the distribution of the latent feature $Z_e$ produced by the VQ-VAE encoder. Finally, the de-normalized latent feature is discretized using the VQ (Sec.\ref{vq}) and passed through the decoder (Sec.\ref{decoder}) to produce the final functional connectivity matrix tasks~$\tilde{X} \in \mathbb{R}^{V \times V}$ for downstream ASD classification tasks.

\subsubsection{Real Sample Guidance} \label{sampling_process}
While the standard reverse diffusion process typically starts from pure Gaussian noise, this approach can sometimes result in out-of-distribution samples. To mitigate this and improve the generation quality, we incorporated ROI features encoded from real connectivity as guidance during the generation process. Inspired by prior work on data augmentation for images \citep{trabucco2023effective, meng2021sdedit}, we spliced an encoded ROI feature from a real connectivity matrix into the diffusion process.

Given a reverse diffusion process with \( T \) steps, we inserted an encoded latent from the real connectivity matrix \( Z_0^{\text{ref}} \) with added noise \(\epsilon \sim \mathcal{N}(0, \mathbf{I})\) at a specific timestep \(\lfloor S t_0 \rfloor\), where guidance level \( t_0 \in [0, 1] \) is a hyperparameter controlling the insertion position:

\begin{align}
Z_{\lfloor S t_0 \rfloor} &= \sqrt{\tilde{\alpha}_{\lfloor S t_0 \rfloor}} Z_0^{\text{ref}} + \sqrt{1 - \tilde{\alpha}_{\lfloor S t_0 \rfloor}} \epsilon \label{eq:real_guidance} \\
S t_0 &= (1 - t_0) \cdot T \label{eq:guidance_level}
\end{align}

The reverse diffusion process then continues from this spliced latent at timestep \(\lfloor S t_0 \rfloor\) until the final sample \( \hat{Z}_g \) is generated. As $t_0 \rightarrow 0$, the generated connectivity are more closely resemble the guidiance connectivity.

\begin{algorithm}[tbp]
\caption{Brain-Net-Diffusion Training}
\label{alg:bnt_train}
\begin{algorithmic}[1]
    \item[\textbf{Input:}] Dataset $\mathcal{D}_R$ consists real functional connectivity $X$ and associated condition $c$
    \item[\textbf{Output:}] Trained Encoder $\mathcal{E}$, Codebook $e$, Decoder $D$ and Noise Predictor $\epsilon_\theta$
    
    \STATE \textbf{initialize} Encoder $\mathcal{E}$, Codebook $e$, Decoder $\mathcal{D}$
    
    \FOR{each training iteration}
        \STATE Connectivity auto-encoding with VQ
        \begin{align*}
            Z_e &\gets \mathcal{E}(X) \\
            Z_q &\gets VQ(Z_e, e) \\
            X' &\gets \mathcal{D}(Z_q)
        \end{align*}
        \STATE Optimize for $\mathcal{E}$ and $e$ with $L_{encode}$ (Eqn. \ref{eq:encode_loss})
    \ENDFOR
    \STATE \textbf{initialize} Diffusion Transformer $\epsilon_\theta$
    
    \FOR{each training iteration}
        \STATE Noise sampling and prediction on normalized latent space
        \begin{align*}
        Z_e &\gets \mathcal{E}(X) \\
        Z_0 &\gets (Z_e - \mu) / \sigma \\
        Z_t &\gets \sqrt{\bar{\alpha}_t} Z_0 + \sqrt{1 - \bar{\alpha}_t} \epsilon, \textit{$\epsilon \sim \mathcal{N}(0, I)$} \\
        \hat{\epsilon} &\gets \epsilon_\theta (Z_t, c, t)
        \end{align*}
        
        \STATE Optimize for $\epsilon_\theta$ with $L_{\epsilon_\theta}$ (Eqn. \ref{eq:diffusion_loss})
    \ENDFOR
\end{algorithmic}
\end{algorithm}

\begin{algorithm}[htbp]
\caption{Brain-Net-Diffusion Generation}
\label{alg:bnt_generate}
\begin{algorithmic}[1]
    \item[\textbf{Requires:}]
    \item[] Trained connectivity encoder $\mathcal{E}$, codebook $e$, decoder $\mathcal{D}$ and noise predictor $\epsilon_\theta$
    \item[]
    
    \STATE \textbf{Function} \textsc{Sampling}($Z_0^{\text{ref}}$, $t0$, $c$):
    \STATE Initialize reverse diffusion process starting point under real sample guidance 
    \begin{align*}
        S t_0 &\gets \lfloor (1 - t_0) \cdot T \rfloor \\
        Z_{S t_0} &\gets \sqrt{\tilde{\alpha}_t} Z_0^{\text{ref}} + \sqrt{1 - \tilde{\alpha}_t} \epsilon\text{, } \epsilon \sim \mathcal{N}(0, \mathbf{I})
    \end{align*}
    \STATE Perform reverse diffusion process from $Z_{S t_0}$ and $S t_0$ to sample a new latent space $\hat{Z_g}$
    \RETURN $\hat{Z_g}$
    \STATE \textbf{End Function}
    \item[]
    
    \STATE \textbf{Function} \textsc{Generate}$(X, c, t0)$:
    \STATE Encode real connectivity to latent space and perform normalization
    \begin{align*}
        Z_e &\gets \mathcal{E}(X) \\
        \hat{Z_e} &\gets (Z_e - \mu) / \sigma \\
    \end{align*}
    
    \STATE Generate a new latent space under the guidance of $\hat{Z_e}$ with the specified condition $c$ and guidance level $t0$
    \begin{align*}
        \hat{Z_g} = \textsc{Sampling}(\hat{Z_e}, t0, c)
    \end{align*}
    
    \STATE De-normalize and decode the generated latent space to functional connectivity
    \begin{align*}
        Z_g &\gets \hat{Z_g} \cdot \sigma + \mu \\
        \tilde{X} &\gets \mathcal{D}(Z_g)
    \end{align*}
    \STATE \textbf{return} $\tilde{X}$
    \STATE \textbf{End Function}
\end{algorithmic}
\end{algorithm}

\begin{algorithm}
\caption{Augmentation Data Construction}
\label{alg:aug_construct}
\begin{algorithmic}
    \item[\textbf{Input:}]
    \item[] (\textit{Data}) Dataset $\mathcal{D}_R$ consists real functional connectivity $X$ and associated condition $c$
    \item[] (\textit{Model}) Brain-Net-Diffusion generation module with \textsc{Generate} that described in Algorithm~\ref{alg:bnt_generate}
    \item[] (\textit{Params}) Guidance level set $t_{\text{guide}}$ and maximum diffusion step $T$
    \item[\textbf{Output:}] Dataset $\mathcal{D}_G$ consists of generated connectivity and their associate condition
    \STATE $\mathcal{D}_G \gets \emptyset$
    
    \FOR{$X$, $c$ in $\mathcal{D}_R$}
        \FOR{$t0$ in $t_{\text{guide}}$}
        \STATE $\tilde{X}_{c} \gets \textsc{Generate}(X, c, t0)$
        \STATE $\tilde{X}_{\neg c} \gets \textsc{Generate}(X, \neg c, t0)$
        \STATE $\mathcal{D}_G \gets \mathcal{D}_G \cup \{\tilde{X}_{c}, \tilde{X}_{\neg c}\}$
        \ENDFOR
    \ENDFOR
\end{algorithmic}
\end{algorithm}

\subsubsection{Condition Contrastive Augmentation} \label{generated_dataset}
For the generation condition $c$ in the reverse diffusion process, with guidance from real samples, $c$ can either match or oppose the condition of the guide real connectivity. We performed generation with both conditions for each guidance real connectivity, to improve the diversity of the generated dataset.

Specifically, we constructed the generated dataset from the original dataset

\begin{equation}
    \mathcal{D}_R = \{(x, c) \mid c = \text{condition of } x\}
\end{equation}

where $x$ is the real functional connectivity for one subject, and $c$ is the condition (ASD or HC) of the subject. 

The generated dataset for augmentation is the union of \textit{guidance identical condition} set $\mathcal{D}_{G(c, t0)}$ and \textit{guidance opposite condition} set $\mathcal{D}_{G(\neg c, t0)}$. The connectivity in $\mathcal{D}_{G(c, t0)}$ is generated with condition identical to the guidance real connectivity:
\begin{equation}
\mathcal{D}_{G(c, t0)} = \{\mathcal{G}(x,c,t0) \mid (x, c) \in \mathcal{D}_R\}
\end{equation}

where $\mathcal{G}$ stands for the trained Brain-Net-Diffusion connectivity generator, $t0$ denotes the guidance level and $c$ denotes the condition (ASD or HC) of the guidance real connectivity $x$. Conversely, $\mathcal{D}_{G(\neg c, t0)}$ consists of connectivity generated under conditions that are opposite to those used in the guidance real connectivity:
\begin{equation}
\mathcal{D}_{G(\neg c, t0)} = \{\mathcal{G}(x,\neg c,t0) \mid (x, c) \in \mathcal{D}_R\}
\end{equation}

The motivation for condition contrastive augmentation with the union of identical and opposite condition-generated data is to maximize the knowledge learned from the conditional latent diffusion and improve the diversity of the generated dataset.

\subsection{Overall Pipeline}
Algorithm~\ref{alg:bnt_train} provides the pseudo code for Brain-Net-Diffusion training process. It includes training for Latent Connectivity Auto-Encoder and Conditional Diffusion Transformer.

Algorithm~\ref{alg:bnt_generate} describes the connectivity generation module for Brain-Net-Diffusion. It integrates the trained connectivity auto-encoder and conditional diffusion transformer to generate new connectivity.

Algorithm~\ref{alg:aug_construct} describes how the augmentation dataset is generated with real sample guidance and condition contrastive augmentation.

\section{Experiment Settings}
\subsection{Dataset and Preprocessing}
We collected rs-fMRI data from 505 ASD and 530 healthy control (HC) subjects from the Autism Brain Imaging Data Exchange I (ABIDE-I) dataset \citep{di2014autism}, which includes data from 17 international sites. This dataset was used to evaluate the performance of our method. Given a series of fMRI readings in $\mathbb{R}^3$ with length $T$, we used the Configurable Pipeline for the Analysis of Connectomes~\citep{li2021moving} to preprocess the fMRI data. The CC200 atlas~\citep{craddock2012whole} was used to segment the fMRI data into ROI signals~$R \in \mathbb{R}^{V \times T}$, where $V$ is the number of ROIs.

The functional connectivity for each subject was constructed by computing the Pearson correlation between all pairs of ROI signals along the time dimension. Formally,

\begin{equation}
    X \in \mathbb{R}^{V \times V}, \, X_{ij} = \mathbf{Pearson}(R_i, R_j)
\end{equation}

where $X$ is the functional connectivity matrix, and $R_i$ and $R_j$ are the signals of the $i$th and $j$th ROI, respectively.

We performed a stratified 5-fold cross-validation based on testing site and diagnosis labels to evaluate the performance of our method. In each fold, 60\% of data was used for training, 20\% for validation, and 20\% for testing.

\subsection{Implementation Details}
Brain-Net-Diffusion was implemented using PyTorch and trained on a Linux system equipped with an NVIDIA RTX A5000 GPU (24GB VRAM) and 256GB of RAM.

The Latent Connectivity Auto-Encoder consists of a linear projection layer followed by four transformer blocks. We set code book size $c = 768$ and codebook vector dimension $k = 16$ . We trained for a maximum of 800 epochs with a batch size of 64, with an average training time of 50 minutes per fold.

The Conditional Diffusion Transformer has 14 DiT blocks and generates a latent vector with a dimension of 128. We set the number of diffusion steps $T$ to 100 and the dimension of time embedding, condition embedding and latent features are all set to 64. We trained for a maximum of 600 epochs with a batch size of 64, with an average training time of 2 hours per fold.

\subsection{Augmentation Settings}
We used the Brain-Net-Diffusion with real sample guidance to augment the training set. We combined dataset generated with Condition Contrastive Augmentation (Sec. 
\ref{generated_dataset}), at guidance level $t0 \in \{0.2, 0.4\}$ (Sec. \ref{sampling_process}), to construct the generated dataset for augmentation.

We adopt a balanced data sampling strategy similar to the approach in \citep{trabucco2023effective} to train the downstream ASD classification model. This strategy ensures each training batch consists of an equal split, with half of the connectivity matrices drawn from the real connectivity set and the other half from the generated connectivity set

We employed a simple MLP classifier with 100-dimensional hidden layer followed by a ReLU activation function and a readout layer to perform downstream ASD classification. The classifier input is the flattened lower triangular portion of the functional connectivity matrix, sized at $\frac{V(V-1)}{2}$. The MLP was trained with a batch size of 256 and a dropout rate of 0.2.

We compared the downstream ASD versus HC classification performance of our methods with two classical augmentation and two generative approaches:

1) \textit{Gaussian Noise} \citep{pei2022data}: This approach generates new samples by adding Gaussian noise to the original connectivity matrix.  It aims to reduce the learning of useless high-frequency features and improve the model robustness. For each 
connectivity matrix, we sampled $N$ Gaussian noises $N(\mu, \sigma)$ with $\mu = 0$ and $\sigma = 0.05$ and added the noise matrix to the original connectivty matrix separately to generate $N$ new samples.

2) \textit{Sliding Windows} \citep{pei2022data}: The sliding window correlation method studies dynamic functional connectivity by continuously sliding and computing pairwise ROI correlations within a window. We adapted the non-overlapping sliding methods, where each ROI signal is equally split into $N$ non-overlap segments along time series and pairwise correlation is computed for each segments.

3) \textit{DRVAE} \citep{qiang2021modeling}: DRVAE combines a VAE with a recurrent neural network to perform data augmentation for time-series data. While the original work is a generative model for ROI signal, we use its generated signal to compute functional connectivity for downstream classification models.

4) \textit{VAE-GAN} \citep{qiang2023functional}: This method combines VAE and GAN to leverage the strengths and mitigate the weaknesses of both models. VAE-GAN collapses the decoder and the generator into one model. We used the generated fMRI time-series data to compute functional connectivity for downstream models.

To ensure a fair and comprehensive comparison, we took account for different augmentation size for each methods. For each original sample, we tested different generated data sizes of 4, 8, 12, 16 for each augmentation method.

\section{Results and Discussion}

\subsection{Overall Performance}
\begin{figure*}[htbp]
    \centering
    \includegraphics[width=0.7\textwidth]{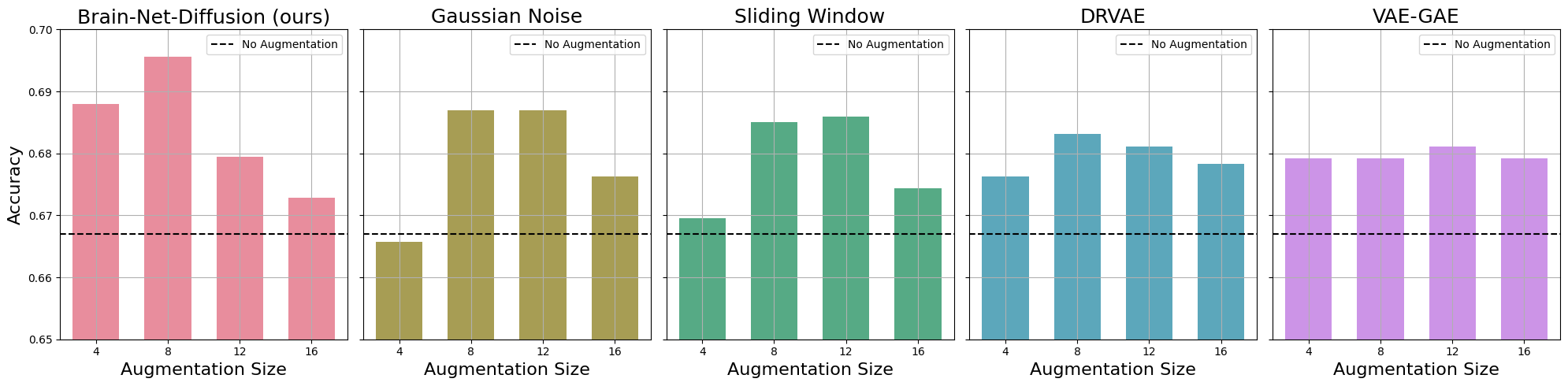}
    \caption{Accuracy for various number of synthetic samples for each augmentation methods.}
    \label{fig:params_study}
\end{figure*}

Figure \ref{fig:params_study} illustrates the classification accuracy comparison across different augmentation methods, varying by the number of augmented samples. Our Brain-Net-Diffusion achieved the highest accuracy at 0.696 with an augmentation size of eight. We observed that both excessively large and very small augmentation sizes did not lead to optimal performance. One possible explanation is that too many generated samples can reduce the overall quality of the dataset, and too few samples do not add enough variety to enhance the augmentation effectively.

Table \ref{tab:aug_comp} reports the ASD classification performance on the test set, comparing performance with and without various augmentation methods. Augmentation using Brain-Net-Diffusion increased accuracy by 4.3\% compared to no augmentation. It also outperformed other augmentation methods by margins ranging from 1.3\% to 2.2\% in accuracy.

\begin{table*}[ht]
\centering
\begin{tabular}{l c c c c}
\toprule
\textbf{Augmentation} & \textbf{AUROC} & \textbf{Accuracy} & \textbf{Sensitivity} & \textbf{Specificity} \\ \midrule

\multirow{1}{*}{No Augmentation} 
& 0.738 $\pm$ 0.030 & 0.667 $\pm$ 0.029 & 0.660 $\pm$ 0.104 & \textbf{0.674 $\pm$ 0.067} \\

\multirow{1}{*}{DRVAE} 
& 0.754 $\pm$ 0.012 & 0.683 $\pm$ 0.006 & 0.697 $\pm$ 0.044 & 0.670 $\pm$ 0.048 \\

\multirow{1}{*}{Gaussian Noise} 
& 0.753 $\pm$ 0.012 & 0.687 $\pm$ 0.014 & 0.713 $\pm$ 0.024 & 0.659 $\pm$ 0.035 \\

\multirow{1}{*}{Sliding Window}
& 0.763 $\pm$ 0.026 & 0.686 $\pm$ 0.027 & 0.714 $\pm$ 0.042 & 0.658 $\pm$ 0.056 \\

\multirow{1}{*}{VAE-GAN} 
& 0.755 $\pm$ 0.012 & 0.681 $\pm$ 0.012 & 0.698 $\pm$ 0.060 & 0.664 $\pm$ 0.048 \\
\midrule

\multirow{1}{*}{BrainNetDiffusion (ours)} 
& \textbf{0.765 $\pm$ 0.021} & \textbf{0.696 $\pm$ 0.012} & \textbf{0.730 $\pm$ 0.042} & 0.660 $\pm$ 0.039 \\
\bottomrule
\end{tabular}
\caption{Comparison of different augmentation methods with their best hyper-parameters}
\label{tab:aug_comp}
\end{table*}

\subsection{Impact of Distribution Normalization and Condition Contrastive Loss}
To validate the effectiveness of our designed modification based on diffusion transformer - distribution normalization and condition contrastive loss, we tested the downstream ASD classification performance with the generated sample from the Brain-Net-Diffusion without each components, respectively. For each real connectivity with guidance level $t0 \in \{0.2, 0.4\}$, we generate two ASD and two HC connectivity matrices, resulting in a total of eight generated samples per real sample.

Table \ref{tab:abl_aug} reports the ASD classification performance on the test set following the removal of Distribution Normalization and Condition Contrastive Loss from our Brain-Net-Diffusion model. The removal of Condition Contrastive Loss alone led to a 2.8\% decrease in accuracy, while the removal of Distribution Normalization resulted in a 3.8\% decrease. When both components were removed simultaneously, accuracy decreased substantially by 9.7\%. This underscores the significant impact of both modules on the augmentation performance of Brain-Net-Diffusion.

\subsection{Impact of Real Sample Guidance}
To validate the necessity of Real Sample Guidance during the generation process (\ref{sampling_process}), we tested the downstream ASD classification performance with the augmented dataset generated by Brain-Net-Diffusion without any guidance and with different guidance rates. We constructed three augmentation set: (1) 4 ASD and 4 HC connectivity with guidance level $t0 = 0.0$ (no guidance), (2) 2 ASD and 2 HC connectivity with guidance level $t0 = \in \{0.2, 0.4\}$ and (3) 2 ASD and 2 HC connectivity with guidance level $t0 = \in \{0.6, 0.8\}$.

Table \ref{tab:abl_guide} illustrates ASD classification performance on the test set with augmented connectivity generated at various guidance levels. Optimal results were achieved with smaller guidance levels of \( t0 \in \{0.2, 0.4\} \). This is expected as generation without guidance may produce low-quality connectivity, and generation with excessive guidance may reduce diversity, rendering the guidance less effective.

\newcommand{\cmark}{\ding{51}}%
\newcommand{\xmark}{\ding{55}}%

\begin{table*}[htbp]
\centering
\begin{tabular}{cccccc}
\toprule
\textbf{DN} & \textbf{CCL} & \textbf{AUROC} & \textbf{Accuracy} & \textbf{Sensitivity} & \textbf{Specificity} \\
\midrule
  &  & 0.711 $\pm$ 0.014 & 0.628 $\pm$ 0.082 & 0.524 $\pm$ 0.297 & 0.739 $\pm$ 0.154 \\
 & \cmark & 0.736 $\pm$ 0.015 & 0.669 $\pm$ 0.024 & 0.714 $\pm$ 0.075 & 0.622 $\pm$ 0.056 \\
\cmark  &  & 0.755 $\pm$ 0.028 & 0.676 $\pm$ 0.025 & 0.687 $\pm$ 0.048 & 0.665 $\pm$ 0.061 \\
\cmark & \cmark & \textbf{0.765 $\pm$ 0.021} & \textbf{0.696 $\pm$ 0.012} & \textbf{0.730 $\pm$ 0.042} & \textbf{0.660 $\pm$ 0.039} \\
\bottomrule
\end{tabular}
\caption{Performance with removal of Distribution Normalization (DN) and Condition Contrastive Loss (CCL)}
\label{tab:abl_aug}
\end{table*}

\begin{table*}[htbp]
\centering
\begin{tabular}{ccccc}
\toprule
\textbf{Guidance Level} & \textbf{AUROC} & \textbf{Accuracy} & \textbf{Sensitivity} & \textbf{Specificity} \\
\midrule
No Guidance & 0.752 $\pm$ 0.020 & 0.678 $\pm$ 0.021 & 0.698 $\pm$ 0.047 & 0.658 $\pm$ 0.030 \\
$t0 \in \{0.2, 0.4\}$ & \textbf{0.765 $\pm$ 0.021} & \textbf{0.696 $\pm$ 0.012} & \textbf{0.730 $\pm$ 0.042} & \textbf{0.660 $\pm$ 0.039} \\
$t0 \in \{0.6, 0.8\}$ & 0.742 $\pm$ 0.023 & 0.673 $\pm$ 0.031 & 0.694 $\pm$ 0.010 & 0.652 $\pm$ 0.056 \\
\bottomrule
\end{tabular}
\caption{Augmentation performance with various guidance level}
\label{tab:abl_guide}
\end{table*}

\begin{table*}[htbp]
\centering
\begin{tabular}{ccccc}
\toprule
\textbf{Condition Set} & \textbf{AUROC} & \textbf{Accuracy} & \textbf{Sensitivity} & \textbf{Specificity} \\
\midrule
$\mathcal{D}_{G(c, t0)}$ (identical) & 0.752 $\pm$ 0.016 & 0.683 $\pm$ 0.017 & 0.719 $\pm$ 0.039 & 0.646 $\pm$ 0.051 \\
$\mathcal{D}_{G(\neg c, t0)}$ (opposite) & 0.725 $\pm$ 0.037 & 0.664 $\pm$ 0.038 & 0.707 $\pm$ 0.062 & 0.618 $\pm$ 0.030 \\
$\mathcal{D}_{G(c, t0)} \cup \mathcal{D}_{G(\neg c, t0)}$ & \textbf{0.765 $\pm$ 0.021} & \textbf{0.696 $\pm$ 0.012} & \textbf{0.730 $\pm$ 0.042} & \textbf{0.660 $\pm$ 0.039} \\
\bottomrule
\end{tabular}
\caption{Augmentation performance with various generation conditions}
\label{tab:abl_cond}
\end{table*}

\subsection{Impact of Condition Contrastive Augmentation}
To validate the necessity of Condition Contrastive Augmentation (\ref{generated_dataset}), we compared the downstream ASD classification performance with (1) the identical condition generated set $\mathcal{D}_{G(c, t0)}$ only, (2) the opposite condition generated set $\mathcal{D}_{G(\neg c, t0)}$ only and (3) the union of (1) and (2). For this experiment, we fixed $t0 \in \{0.2, 0.4\}$. For the identical and opposite generated set, we generated four connectivity for each corresponding real guidance. For the union, we generated two connectivity for each condition for each corresponding real guidance. Thus, augmentation size for each real connectivity is fixed at eight for all generated set in this experiment.

Table \ref{tab:abl_cond} shows the ASD classification performance on the test set with augmentation consisting of identical, opposite, and union conditional generations. In terms of AUROC and accuracy, augmentation using identical conditions proved less effective than the other two approaches. The union of identical and opposite condition augmentation achieved optimal performance, demonstrating the benefit of diversified training conditions in enhancing model robustness and generalization.

\subsection{Evaluation for Generated Connectivity}
We used qualitative and quantitative evaluation methods to interpret the generated functional connectivity matrices and evaluate their impacts on improving the performance of the downstream ASD classification model.

\subsubsection{Quantitative Evaluation}
We computed the average pairwise MAE to quantify the differences between the real and generated functional connectivity matrices. Specifically, each entry in the average pairwise MAE is defined by

\begin{equation}
    \mathbf{MAE}_{ij} = \frac{1}{\vert\mathcal{D}\vert} \sum_{x \in \mathcal{D}} MAE(x_{i}, x_{j})
\end{equation}

where $\mathcal{D}$ is the dataset of all subjects' connectivity, $x_{i}$ is real connectivity and $x_{j}$ is the corresponding generated connectivity matrix, guided by $x_{i}$ at different guidance levels. We generate one connectivity for each real connectivity at each guidance level of 0.0, 0.2, 0.4, 0.6 and 0.8.

\begin{figure*}
    \centering
    \begin{subfigure}{0.3\textwidth}
        \centering
        \includegraphics[width=\textwidth]{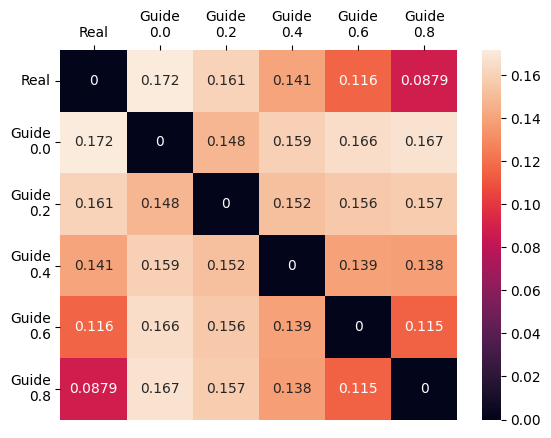}
        \caption{Training Set Pairwise MAE}
    \end{subfigure}
    \begin{subfigure}{0.3\textwidth}
        \centering
        \includegraphics[width=\textwidth]{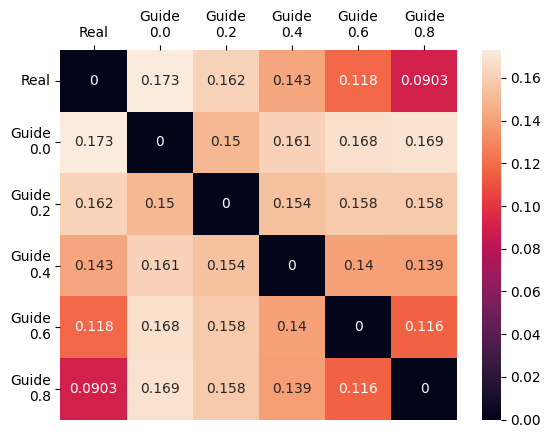}
        \caption{Testing Set Pairwise MAE}
    \end{subfigure}
    \caption{Pairwise Mean-Absolute-Error (MAE) between all real and generated functional connectivity}
    \label{fig:pair_mae}
\end{figure*}

Figure \ref{fig:pair_mae} illustrates the pairwise Mean Absolute Error (MAE) between real and generated connectivity. The results reveal three key observations. First, the pairwise MAE for train and test generation results are similar, indicating that Brain-Net-Diffusion demonstrates strong generalization capabilities. Second, the MAE between real and generated connectivity decreases as the guidance level increases, aligning with the expectation that higher guidance levels produce connectivity that more closely resembles the real data. Lastly, the MAE between generated connectivity and connectivity generated at different guidance levels is inversely proportional to the target connectivity guidance level. Specifically, connectivity generated with a lower guidance level exhibits a larger MAE when compared to connectivity generated with a higher guidance level.

\begin{figure*}
    \centering
    \begin{subfigure}[t]{0.1\textwidth}
        \centering
        \includegraphics[width=\textwidth]{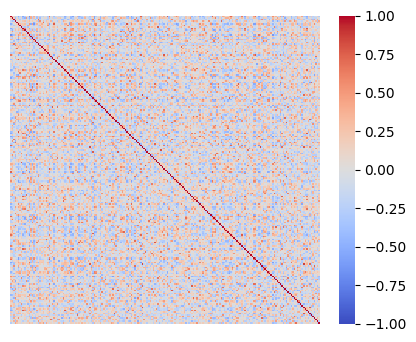}
        \label{fc_visual_real}
    \end{subfigure}
    \begin{subfigure}[t]{0.6\textwidth}
        \centering
        \includegraphics[width=\textwidth]{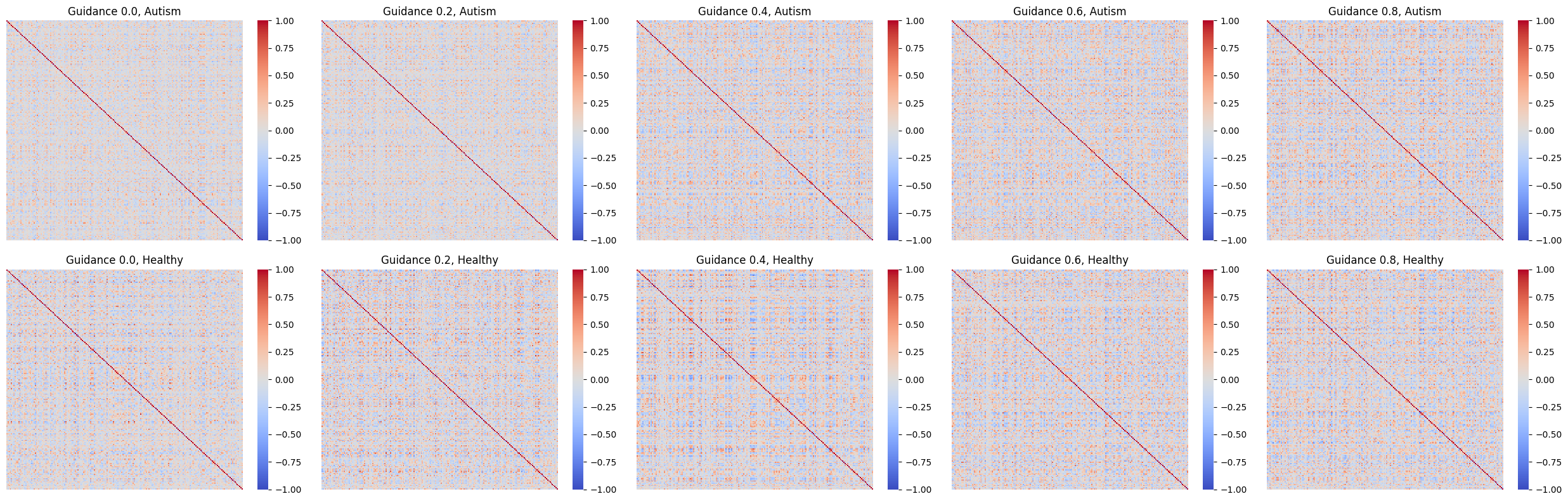}
        \label{fc_visual}
    \end{subfigure}
    \caption{Sample 50003 (Autism) real (left) and corresponding generated (right) connectivity.}
    \label{fig:fc_vis}
\end{figure*}

\begin{figure*}
    \centering
    \begin{subfigure}{0.6\textwidth}
        \centering
        \includegraphics[width=\textwidth]{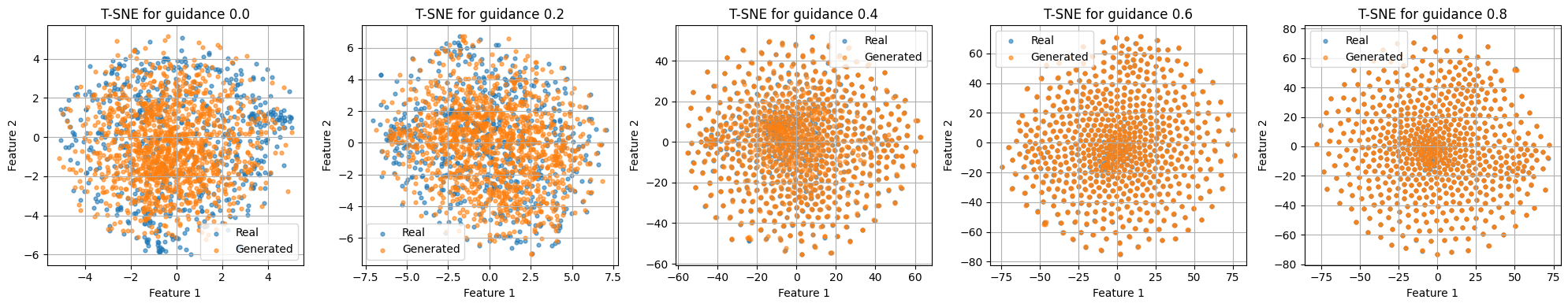}
    \end{subfigure}
    \hfill
    \begin{subfigure}{0.6\textwidth}
        \centering
        \includegraphics[width=\textwidth]{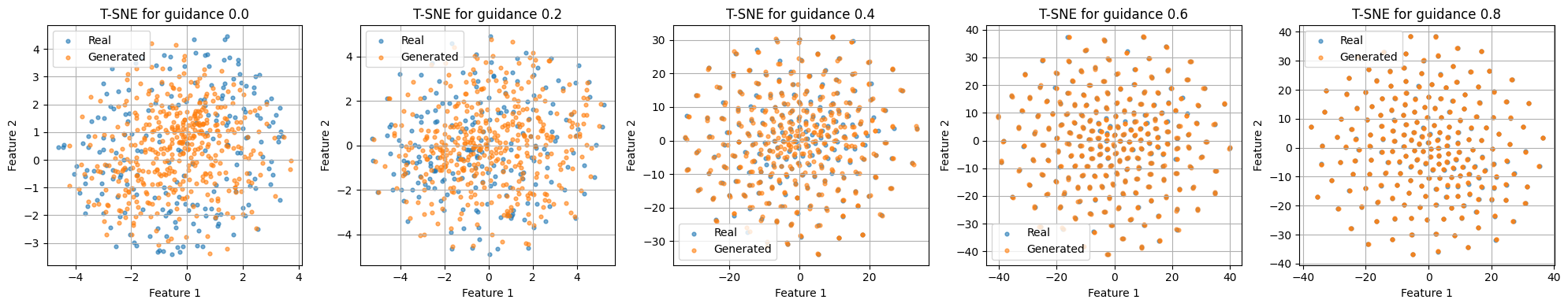}
    \end{subfigure}
    \caption{T-SNE for real and generated connectivity, train (top) and test (bottom) of fold-0}
    \label{fig:tsne}
\end{figure*}

\subsubsection{Qualitative Evaluation}
We visualized the real and generated functional connectivity
matrices to qualitatively evaluate the structure and realism of the generated matrices. First, we draw the adjacency matrix of both real and generated connectivity to visually compare their structures. Next, we flattened the adjacency matrices and applied T-SNE dimensionality reduction \citep{van2008visualizing} to project the high-dimensional connectivity into a lower-dimensional space. It allowed us to observe clustering patterns and assess how well the generated matrices align with the real connectivity. We generated one ASD and one HC functional connectivity with guidance level 0.2, 0.4, 0.6 and 0.8 for samples in both train and test set.

Figure \ref{fig:fc_vis} presents a real functional connectivity alongside its generated counterparts produced by Brain-Net-Diffusion at varying guidance levels and conditions. The results show that the generated connectivity networks are visually similar to the real ones. This visual similarity provides a preliminary indication that Brain-Net-Diffusion is capable of generating functional connectivity that closely resembles real data.

Figure~\ref{fig:tsne} presents the T-SNE visualization of real and generated functional connectivity, of training and testing set, respectively. The results show that the generated connectivity follows a similar distribution with the real connectivity, and the similarity increases as the guidance level increases.

\subsection{Biomarker Interpretations}
We designed and validated a sampling and distance computation algorithm to identify the important ROIs for ASD classification, based on our trained Brain-Net-Diffusion. Furthermore, by mapping the ROIs to the Yeo 7 Networks \citep{yeo2011organization}, we provided a function module-level analysis for ASD brain dysfunction.

\begin{figure*}
    \centering
    \begin{minipage}{0.8\textwidth}
        \centering
        \begin{subfigure}{0.48\textwidth}
            \centering
            \includegraphics[width=\textwidth]{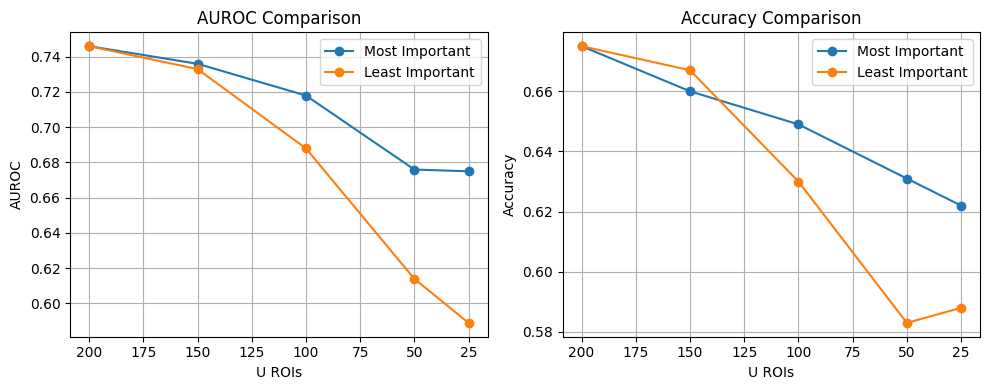}
            \caption{Comparison between classification performance with most and least important ROIs}
            \label{fig:top_roi_val}
        \end{subfigure}
        \hfill
        \begin{subfigure}{0.48\textwidth}
            \centering
            \includegraphics[width=\textwidth]{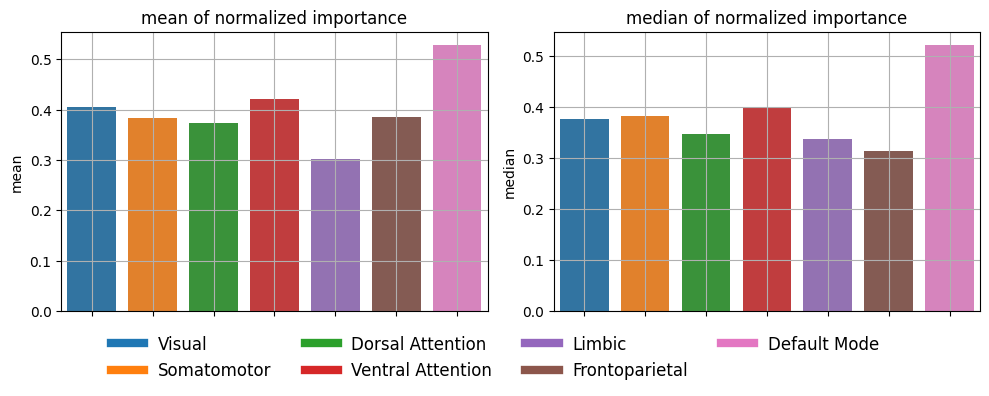}
            \caption{Mean and median importance score for each of the Yeo 7 Networks.}
            \label{fig:yeo_score_agg}
        \end{subfigure}
    \end{minipage}
    \vspace{0.5em} 
    \begin{minipage}{0.6\textwidth}
        \centering
        \includegraphics[width=\textwidth]{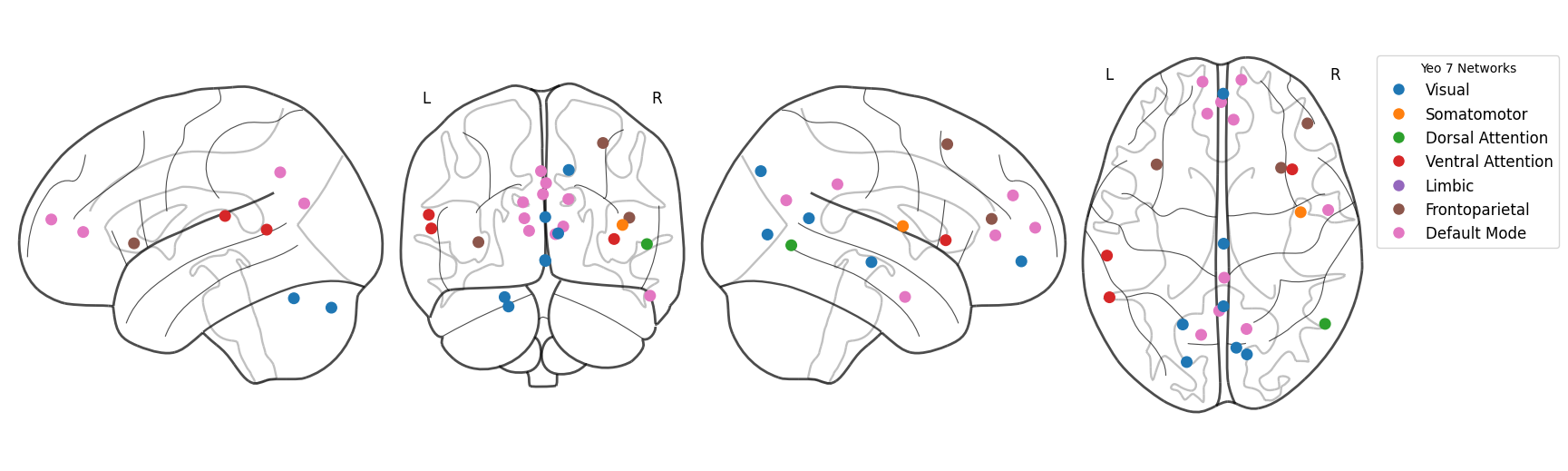}
        \caption{Center of mass of the most important 25 ROIs}
        \label{fig:roi_marker}
    \end{minipage}
    \label{fig:combined_figures}
\end{figure*}

\subsubsection{ROI importance computation}
To analysis the importance of each ROI for ASD classifications, we leveraged the trained noise predictor $\epsilon_\theta$ within the Conditional Diffusion Transformer (Sec. \ref{reverse_diffusion}). Starting from the same initial Gaussian noise, the noise predictor generates different predicted noise for each ROI under different conditions (ASD and HC). This variation arises due to adaptive layer normalization (Equation~\ref{eq:adaln}), which scales and shifts the features of each ROI according to the given condition. By measuring the differences in predicted noise between the two conditions for each ROI, we assess the sensitivity of each ROI to the condition change. The greater the difference, the more sensitive the ROI is, indicating its potential importance in classification.

Algoithm~\ref{alg:roi_importance} presents the ROI importance score calculation method. It first samples Gaussian noise and generates predicted noise for both conditions (ASD and HC) using the noise predictor $\epsilon_\theta$. For each ROI,  we calculated the Euclidean distance between the predicted noise under the two conditions. These distances are averaged over multiple noise samples to produce a final importance score for each ROI, where higher scores indicate greater sensitivity and thus higher importance for classification.

\begin{algorithm}[htbp]
\caption{Compute Importance for Each ROI}
\label{alg:roi_importance}
\begin{algorithmic}[1]
\item[\textbf{Input:}] sampling size $M$, trained diffusion transformer noise predictor $\epsilon_\theta$, maximum scheduled diffusion step $t_{max}$, number of ROI (nodes) $R$, condition label for ASD $c_{ASD}$ and condition label for HC $c_{HC}$.
\item[\textbf{Output:}] importance score for each ROI
\item[]

\STATE $D \gets \mathbf{0} \in \mathbb{R}^{M \times R}$
\FOR{$i \gets 1$ to $M$}
    \STATE $\mathbf{z} \sim \mathcal{N}(\mathbf{0}, \mathbf{I})$
    \STATE $\epsilon_{HC} \gets \epsilon_\theta(\mathbf{z}, t_{max}, c_{h})$
    \STATE $\epsilon_{ASD} \gets \epsilon_\theta(\mathbf{z}, t_{max}, c_{a})$
    \FOR{$r \gets 1$ to $R$}
        \STATE $D_{i,r} \gets \| \epsilon_{HC, r} - \epsilon_{ASD, r}\|^2_{2}$
    \ENDFOR
\ENDFOR
\item[]

\STATE $\mathbf{d} \gets \mathbf{0} \in \mathbb{R}^{R}$
\FOR{$r \gets 1$ to $R$}
    \STATE $d_{r} \gets \frac{1}{M} \sum_{i = i}^{M} D_{i,r}$
\ENDFOR

\STATE \textbf{return} $\mathbf{d}$
\label{node_select}
\end{algorithmic}
\end{algorithm}

\subsubsection{Importance Score Validation}
To validate the importance of ROIs interpreted from the Brain-Net-Diffusion, we compared the ASD classification performance on the functional connectivity of the $U$ most important ROIs versus the $U$ least important ROIs. The hypothesis is that if ROIs are correctly ranked by their importance, as defined by Brain-Net-Diffusion, then the most important ROIs should capture more ASD-relevant features, leading to better classification performance compared to the least important ROIs.

We started with the original functional connectivity matrix $X \in \mathbb{R}^{V \times V}$ and constructed a new connectivity matrix of shape $U \times U$ using the rows and columns corresponding to the selected $U$ ROIs. An MLP classifier with 100 hidden units was trained using the flattened lower triangular part of this cropped connectivity matrix as input.

Figure \ref{fig:top_roi_val} reports the MLP classification performance on the connectivity of the top-$U$ most important and the  bottom-$U$ least important ROIs, $U \in \{200,150,100,50,25\}$. The results reveal that while classification performance declines for both sets of ROIs as $U$ decreases, the ROIs identified as important by the diffusion model consistently outperform those deemed least important. This supports the accuracy of the importance scores assigned to the ROIs by the diffusion model.

\subsubsection{Functional Module Analysis}
We provided a functional module-level analysis by mapping the ROIs and their importance scores to Yeo 7 networks \citep{yeo2011organization}. Yeo 7 networks divide the brain into seven distinct functional networks based on resting-state fMRI, including Visual, Somatomotor, Dorsal Attention, Ventral Attention, Limbic, Frontoparietal and Default Mode. 

Figure \ref{fig:roi_marker} shows the center of mass of the 25 most important ROIs interpreted by our diffusion transformer, along with their respective networks. These ASD-relevant ROIs are distributed across various brain regions, with 7 ROIs belonging to the Visual network and 10 ROIs to the Default Mode network.

We calculated the average and median importance scores for the ROIs mapped into each network as a measurement to evaluate which network is most important for ASD classification,. Figure \ref{fig:yeo_score_agg} illustrates these scores for all 7 networks. The Default Mode network shows the highest mean and median importance scores. In contrast, the Limbic network has the lowest mean importance score, and the Frontoparietal network exhibits the lowest median importance score.

These results suggest that the Default Mode network is most strongly related to ASD classification. This aligns with behavioral observations from fMRI studies and supports the prevailing theory that ASD involves alterations in cognitive networks while preserving or altering sensory processing areas\citep{robertson2013tunnel,turkeltaub2004neural,iuculano2014brain}.

\section{Conclusion}
In conclusion, we propose Brain-Net-Diffusion to address the issue of data scarcity in fMRI-based ASD classification. By augmenting the training data using the Brain-Net-Diffusion model, the downstream ASD classification model can leverage a richer dataset to improve accuracy in identifying ASD. Additionally, we introduce two new modules in the training and generation process: distribution normalization and conditional contrastive loss, which contribute to a more sophisticated training approach. We conducted extensive experiments and analysis. Our results indicate that our proposed data augmentation model can significantly enhance the performance of ASD classification model. Future research could explore the application of Brain-Net-Diffusion to other neurodevelopmental disorders and investigate its potential integration with other advanced machine learning techniques.

\begingroup
\fontsize{8}{10}\selectfont
\bibliographystyle{dcu}
\bibliography{reference}
\endgroup

\end{document}